\documentclass[12pt,english,longbibliography,nofootinbib,superscriptaddress,12pt,sort&compress,showkeys]{revtex4}
\usepackage[T1]{fontenc}
\usepackage[latin9]{inputenc}
\usepackage{xcolor}
\usepackage{array}
\usepackage{textcomp}
\usepackage{multirow}
\usepackage{amsmath}
\usepackage{amssymb}
\usepackage{graphicx}
\usepackage{esint}
\PassOptionsToPackage{normalem}{ulem}
\usepackage{ulem}

\makeatletter

\newcommand{\lyxmathsym}[1]{\ifmmode\begingroup\def\b@ld{bold}
  \text{\ifx\math@version\b@ld\bfseries\fi#1}\endgroup\else#1\fi}

\providecommand{\tabularnewline}{\\}
\providecolor{lyxadded}{rgb}{0,0,1}
\providecolor{lyxdeleted}{rgb}{1,0,0}
\DeclareRobustCommand{\lyxsout}[1]{\ifx\\#1\else\sout{#1}\fi}

\@ifundefined{textcolor}{}
{%
 \definecolor{BLACK}{gray}{0}
 \definecolor{WHITE}{gray}{1}
 \definecolor{RED}{rgb}{1,0,0}
 \definecolor{GREEN}{rgb}{0,1,0}
 \definecolor{BLUE}{rgb}{0,0,1}
 \definecolor{CYAN}{cmyk}{1,0,0,0}
 \definecolor{MAGENTA}{cmyk}{0,1,0,0}
 \definecolor{YELLOW}{cmyk}{0,0,1,0}
}

\usepackage{indentfirst}%\DeclareGraphicsExtensions{.pdf,.png,.jpg}
\usepackage{amsfonts}
\usepackage[T1]{fontenc}
\usepackage{ae,aecompl}
\usepackage{sidecap}
\usepackage[section]{placeins}
\usepackage{epsf}
\makeatother

\usepackage{babel}
\date{\today}

\usepackage{chngcntr}

\makeatother

\usepackage{babel}
\begin{document}
\title{Atomistic study of grain-boundary segregation and grain-boundary diffusion
in Al-Mg alloys }
\author{\noindent R. K. Koju}
\address{Department of Physics and Astronomy, MSN 3F3, George Mason University,
Fairfax, Virginia 22030, USA}
\author{\noindent Y. Mishin}
\address{Department of Physics and Astronomy, MSN 3F3, George Mason University,
Fairfax, Virginia 22030, USA}
\begin{abstract}
Mg grain boundary (GB) segregation and GB diffusion can impact the
processing and properties of Al-Mg alloys. Yet, Mg GB diffusion in
Al has not been measured experimentally or predicted by simulations.
We apply atomistic computer simulations to predict the amount and
the free energy of Mg GB segregation, and the impact of segregation
on GB diffusion of both alloy components. At low temperatures, Mg
atoms segregated to a tilt GB form clusters with highly anisotropic
shapes. Mg diffuses in Al GBs slower than Al itself, and both components
diffuse slowly in comparison with Al GB self-diffusion. Thus, Mg segregation
significantly reduces the rate of mass transport along GBs in Al-Mg
alloys. The reduced atomic mobility can be responsible for the improved
stability of the microstructure at elevated temperatures.
\end{abstract}
\maketitle

\section{Introduction}

Al-Mg alloys constitute an important class of lightweight structural
materials that find numerous automotive, marine and military applications
\citep{Polmear:2017}. Mg improves many mechanical properties of Al,
such as tensile and fatigue strength, ductility, and weldability \citep{Polmear:2017,Lathabai:2002,Pun:2017aa,Devaraj:2019aa},
while maintaining a high strength to weight ratio and a relatively
low production cost. Progress in designing more advanced Al-Mg alloys
requires further improvements in the fundamental knowledge of the
Mg effect on the microstructure and properties. 

Previous experimental and modeling studies have shown that Mg segregates
to Al grain boundaries (GBs), modifying their thermodynamic and kinetic
properties \citep{Guillot:2010aa,Liu:1998aa,Liu:2005aa,Lee:2016aa,Pun:2017aa,Devaraj:2019aa,Rahman:2016aa,Kazemi:2019aa}.
Mg segregation was found to increase both the strength and ductility
of Al, as well as thermal stability of the grains \citep{Lee:2016aa,Pun:2017aa,Devaraj:2019aa,Kazemi:2019aa}.
The stability improvement is attributed to a combination of the thermodynamic
reduction in the GB free energy and the pinning of GBs by solute atoms
due to the solute drag effect. It should be emphasized that the solute
drag process is controlled by diffusion of the solute atoms in the
GB region \citep{Cahn-1962,Lucke-Stuwe-1963,Lucke:1971aa,Mishin:2019aa,Koju:2020aa}.
Diffusion must be fast enough to move the segregation atmosphere along
with the moving boundary. If diffusion is too slow and/or the GB motion
too fast, the boundary breaks away from the segregation atmosphere
and the drag force abruptly drops \citep{Cahn-1962,Mishin:2019aa}.
On the other hand, fast GB diffusion promotes coarsening of the microstructure
by accelerating the mass transport of the alloy components. A detailed
understanding of the GB diffusion process and its relationship with
solute segregation is a prerequisite for rational design of Al-Mg
alloys.

When the Al matrix is supersaturated with Mg, the excess Mg atoms
diffuse toward and then along GBs and precipitate in the form of the
Al$_{3}$Mg$_{2}$ phase and/or possibly other, metastable compounds
\citep{Devaraj:2019aa,Birbilis:2005,Yan:2016aa}. Such precipitates
usually have a detrimental effect by causing, for example, corrosion
cracking and other undesirable consequences \citep{Searles:2001aa}.
The GB precipitation process depends on the level of GB segregation
and the rate of Mg GB diffusion. 

Surprisingly, while Al GB diffusion in Mg has been measured \citep{Das:2014aa,Zhong:2020aa},
to the best of our knowledge, Mg GB diffusion coefficients in Al or
Al-Mg alloys have not been measured experimentally or predicted by
simulations. The only paper known to us \citep{Gangulee75} contains
highly indirect estimates of the triple product $s\delta D$ ($s$
being a segregation parameter, $\delta$ the GB width, and $D$ the
GB diffusion coefficient)\footnote{The units of $s\delta D$ were not given in \citep{Gangulee75}, but
it was later suggested \citep{Kaur89}, based on previous papers of
these authors, that they could be cm$^{3}$\,s$^{-1}$.} based on electromigration experiments in thin films at one temperature.
These measurements do not provide a complete or reliable quantitative
information on Mg GB diffusion coefficients.

In this paper, we report on detailed atomistic computer simulations
of GB segregation and GB diffusion in the Al-Mg system, focusing on
a particular Al-5.5at.\%Mg composition relevant to industrial alloys.
Two representative GBs were selected, a high-angle tilt GB composed
of closely spaced structural units, and a low-angle twist GB composed
of discrete dislocations. The latter case essentially probes the dislocation
segregation effect and the dislocation pipe diffusion. In addition
to computing some of the key characteristics of Mg GB segregation
over a range of temperatures, the simulations reveal some interesting
features of the segregation, such as the formation of Mg clusters
in the high-angle GB and the tendency of the clusters to have highly
elongated shapes reminiscent of linear atomic chains. The diffusion
coefficients and Arrhenius parameters have been computed for GB diffusion
of both Al and Mg, and are compared with Al GB self-diffusion as well
as diffusion of both components in liquid alloys.

\section{Methodology\label{sec:Methodology}}

Atomic interactions in the Al-Mg system were modeled using the Finnis-Sinclair
potential developed by Mendelev et al.~\citep{Mendelev:2009aa}.
The potential provides an accurate description of the Al-rich part
of the phase diagram and predicts the melting temperatures of Al and
Mg to be 926 K \citep{Mendelev08a} and 914 K \citep{Mendelev:2006aa},
respectively, in good agreement with experimental data (934 and 922
K, respectively). The software package LAMMPS (Large-scale Atomic/Molecular
Massively Parallel Simulator) \citep{Plimpton95} was utilized to
conduct molecular statics, molecular dynamics (MD), and Monte Carlo
(MC) simulations. Visualization and structural analysis were performed
using the OVITO software \citep{Stukowski2010a}.

The high-angle GB studied here was the symmetrical tilt $\Sigma17(530)[001]$
GB with the misorientation angle of $61.93^{\circ}$. The parameter
$\Sigma$ is the reciprocal density of coincident sites, {[}001{]}
is the tilt axis, and $(530)$ is the GB plane. This boundary was
created by aligning the crystallographic plane $(530)$ parallel to
the $x$-$y$ plane of the Cartesian coordinate system and rotating
the upper half of the simulation block ($z>0$) by $180^{\circ}$
about the $z$-axis. The low-angle GB was the $\Sigma3601(001)$ twist
boundary with the misorientation angle of $1.91^{\circ}$. In this
case, the GB plane is $(001)$ and the two lattices are rotated relative
to each other about the common $[001]$ axis. The simulation blocks
had approximately square cross-sections parallel to the GB plane.
The block dimensions in the $x$, $y$ and $z$ directions were, respectively,
$11.79\times11.73\times23.67$ nm ($1.97\times10^{5}$ atoms) for
the high-angle GB and $24.27\times24.27\times48.56$ nm ($1.72\times10^{6}$
atoms) for the low-angle GB. Periodic boundary conditions were imposed
in all three directions.

The initial GB structures were optimized by the $\gamma$-surface
method \citep{Mishin98a,Suzuki03a,Suzuki03b}. In this method, one
grain is translated relative to the other by small increments parallel
to the GB plane. After each increment, the total energy is minimized
with respect to local atomic displacements and rigid translations
of the grains normal to the GB plane (but not parallel to it). The
minimized GB energy is plotted as a function of the translation vector,
producing a so-called $\gamma$-surface. The translation vector corresponding
to the deepest energy minimum on the $\gamma$-surface is identified,
and the total energy is further minimized by allowing arbitrary atomic
displacements in all three directions staring from this translational
state. The GB structure obtained is considered the closest approximation
of the ground state of the boundary.

To create a thermodynamically equilibrium distribution of Mg atoms
in the Al-5.5at.\%Mg alloy, the hybrid MC/MD algorithm \citep{Sadigh2012}
was implemented in the semi-grand canonical NPT ensemble (fixed total
number of atoms $N$, fixed temperature $T$, and zero pressure $P$).
Every MC step was followed by 250 MD steps with the integration time
step of 2 fs. The imposed chemical potential difference between Al
and Mg was adjusted to produce the desired chemical composition inside
the grains. The simulation temperature varied between 350 K and 926
K. 

GB diffusion was studied by NPT MD simulations in the temperature
range from 400 K to 926 K using the GBs pre-equilibrated by the MC/MD
procedure. During the MD runs, the GB position could slightly vary
due to thermal fluctuations. To account for such variations, the instantaneous
GB position was tracked by finding the peak of the potential energy
(averaged over thin layers parallel to the GB plane) as a function
the $z$ coordinate normal to the boundary. The GB position was identified
with the center of the peak, while the GB width $\delta$ was estimated
from the peak width. Based on these estimates, the GB core region
was defined as the layer centered at the peak and having the width
of $\delta=1$ nm for the high-angle GB and $\delta=1.5$ nm for the
low-angle GB. The mean-square displacements, $\left\langle x^{2}\right\rangle $
and $\left\langle y^{2}\right\rangle $, of both Al and Mg atoms parallel
to the GB plane were computed as functions of time. The calculations
extended over a time period $\Delta t$ ranging from 0.03 ns to 120
ns, depending on the temperature. The GB diffusion coefficients of
both species were extracted from the Einstein relations $D_{x}=\left\langle x^{2}\right\rangle /2\Delta t$
and $D_{y}=\left\langle y^{2}\right\rangle /2\Delta t$, respectively.
For comparison, similar calculations here performed for Al self-diffusion
in both GBs. In this case, the pure Al boundary was equilibrated by
a 2 ns MD run before computing the mean-square displacements. For
the low-angle GB, the symmetry dictates that $D_{x}$ and $D_{y}$
must be equal. Accordingly, the diffusion coefficients reported for
this boundary were averaged over both directions.

For further comparison, the same methodology was applied to compute
the diffusion coefficients of Al and Mg in the liquid Al-5.5at.\%Mg
alloy at temperatures close to the solid-liquid coexistence (solidus)
line. The simulation block had the dimensions of $11.73\times11.73\times11.73$
nm ($\sim10^{5}$ atoms) and was equilibrated by an MD run for a few
ns prior to diffusion calculations.

\section{Results and analysis}

\subsection{Grain boundary structures and energies}

\begin{figure}[ht]
\textbf{(a)} \includegraphics[height=0.29\textwidth]{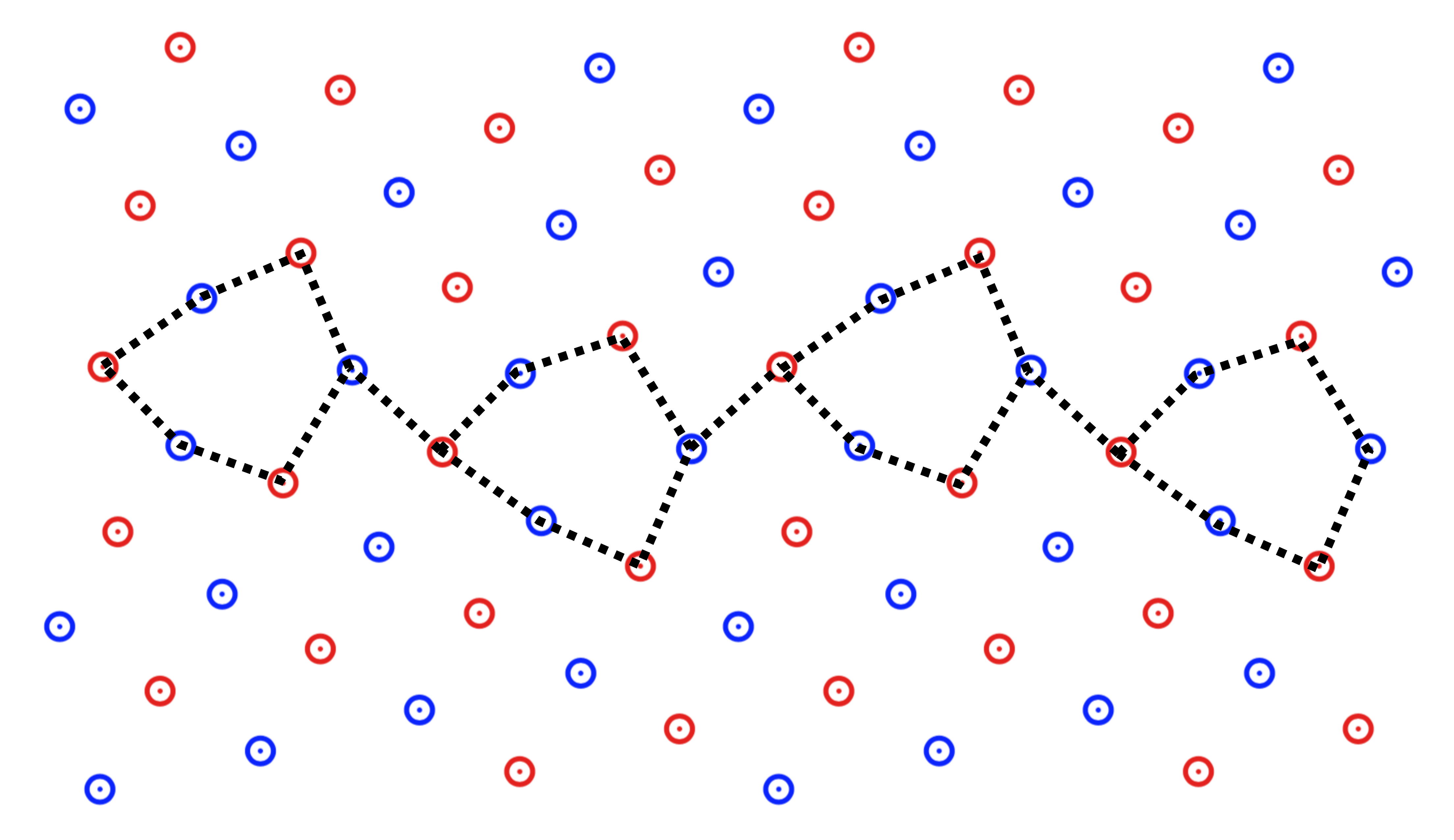} 

\bigskip{}

\textbf{(b)} \includegraphics[height=0.33\textwidth]{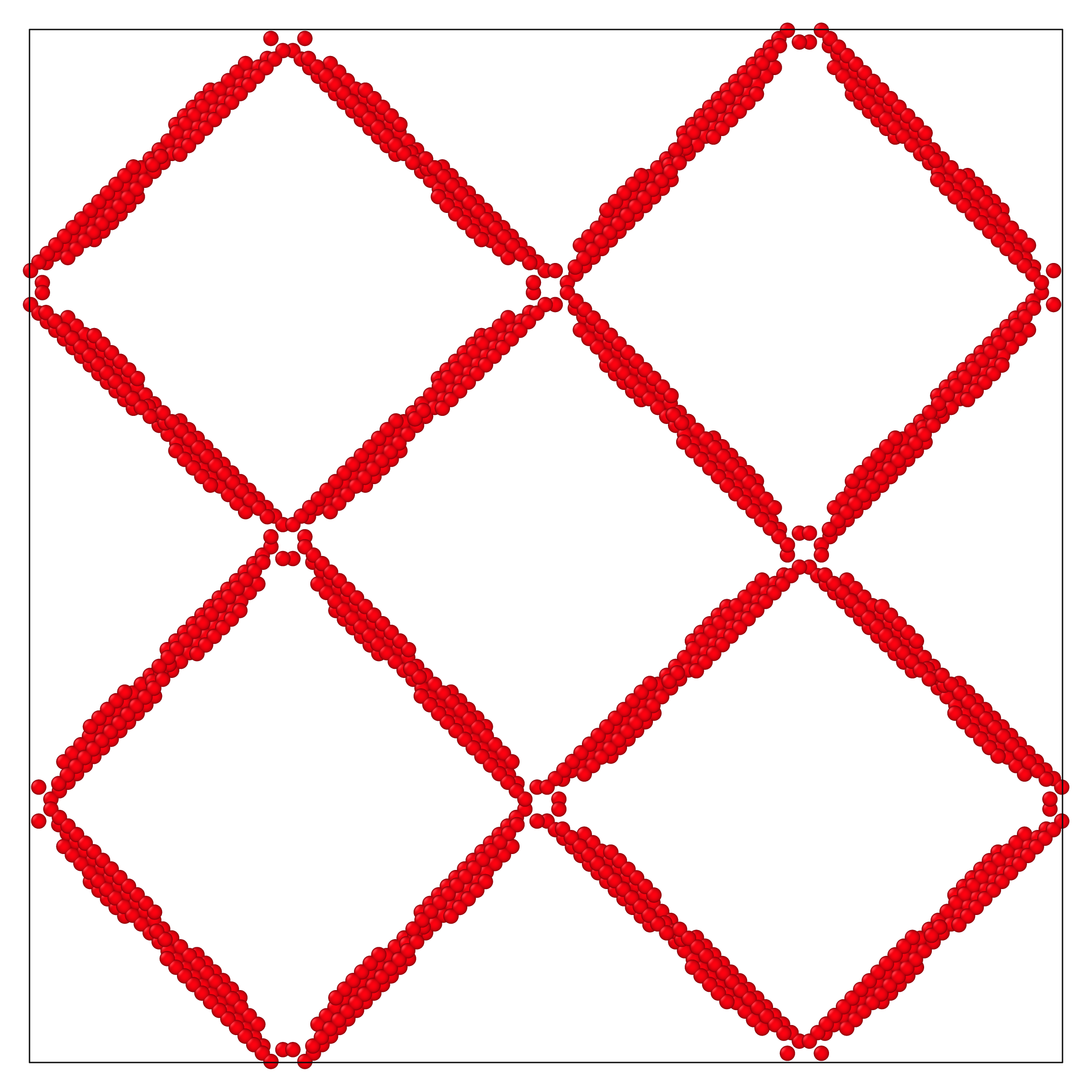} \caption{Structures of the GBs studied in this work. (a) Symmetrical tilt $\Sigma17(530)[001]$
GB composed of kite-shape structural units. The structure is projected
along the {[}001{]} tilt axis normal to the page. The GB plane is
horizontal. The open and filled circles represent atoms located in
alternating (002) planes parallel to the page. The structural units
are outlined by dotted lines. (b) Top view of the $\Sigma3601(001)$
twist GB composed of $\tfrac{1}{2}\left\langle 110\right\rangle $
edge dislocations. The $\{001\}$ GB plane is parallel to the page.
The dislocations are visualized by the bond-order analysis using OVITO
\citep{Stukowski2010a}. The perfect-lattice atoms are removed for
clarity.}
\label{fig:GB17530}
\end{figure}

The excess energy of the equilibrated high-angle $\Sigma17$ GB was
found to be 488 mJ\,m$^{-2}$. The 0 K structure of this boundary
consists of identical kite-shape structural units arranged in a zigzag
array as shown in Fig.~\ref{fig:GB17530}a. The rows of these structural
units running parallel to the tilt axis (normal to the page) can be
interpreted as an array of closely spaced edge dislocations forming
the GB core. An identical zigzag arrangement of the kite-shape units
in this GB was earlier found in Cu \citep{Suzuki05a,Cahn06b,PhysRevMaterials.4.073403,Koju:2020aa}
and Ni \citep{Sun:2020aa}, suggesting that this atomic structure
is common to FCC metals.

The low-angle $\Sigma3601$ GB has a smaller energy of 127 mJ\,m$^{-2}$
and consists of a square network of discrete dislocations (Fig.~\ref{fig:GB17530}b).
As expected from the dislocation theory of GBs \citep{Hirth}, the
dislocation lines are parallel to the $\left\langle 110\right\rangle $
directions and have the Burgers vectors of $\mathbf{b}=\tfrac{1}{2}\left\langle 110\right\rangle $.
Furthermore, the Frank formula \citep{Hirth} predicts that the distance
between parallel GB dislocations in the network must be approximately
$|\mathbf{b}|/\theta$, where $\theta$ is the twist angle. Examination
of the GB structure reveals that this prediction is indeed followed
very closely. 

\subsection{Grain boundary segregation}

\begin{figure}[ht]
\textbf{(a)} \includegraphics[width=0.35\textwidth]{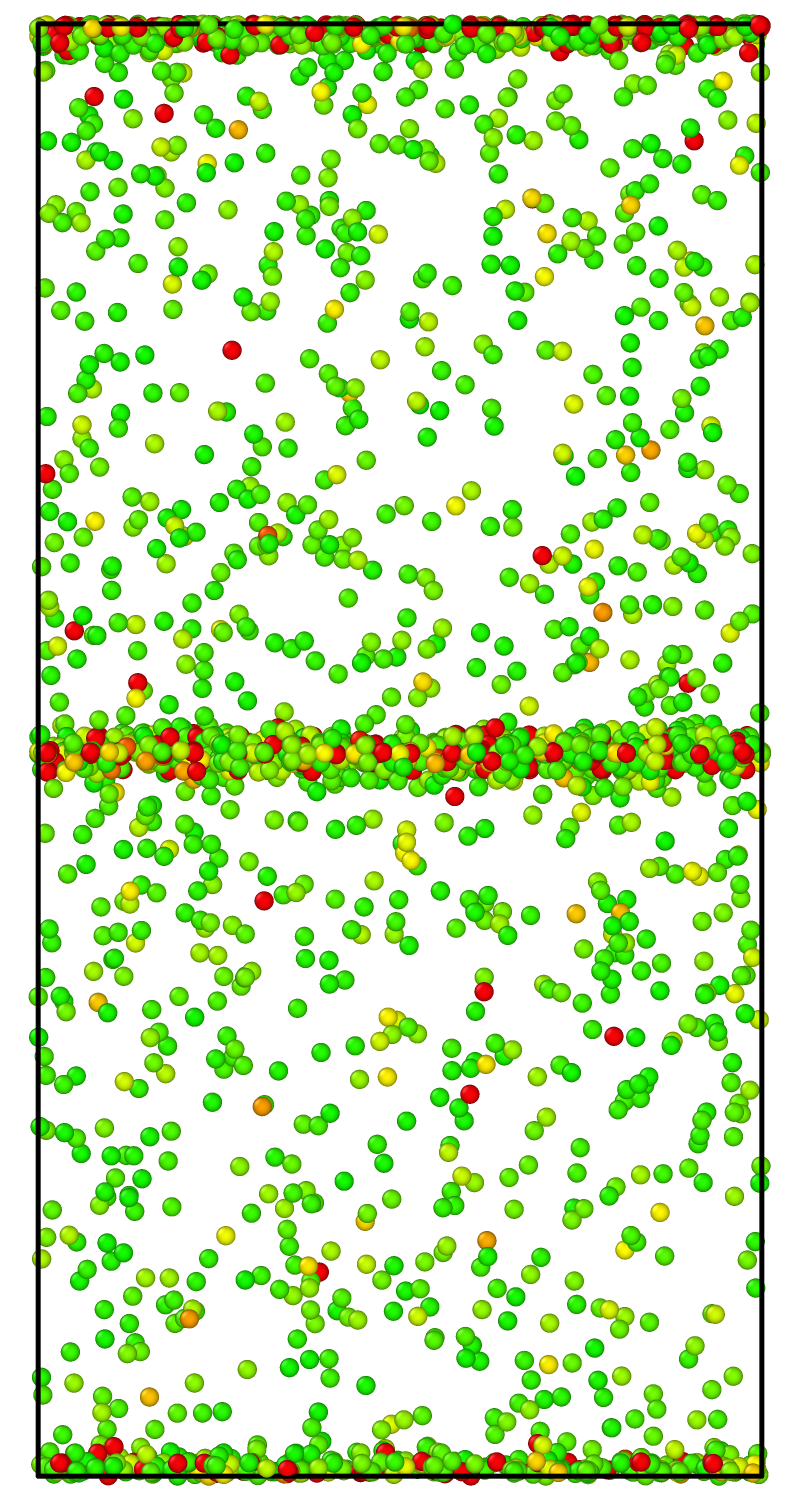}

\bigskip{}

\textbf{(b)} \includegraphics[width=0.5\columnwidth]{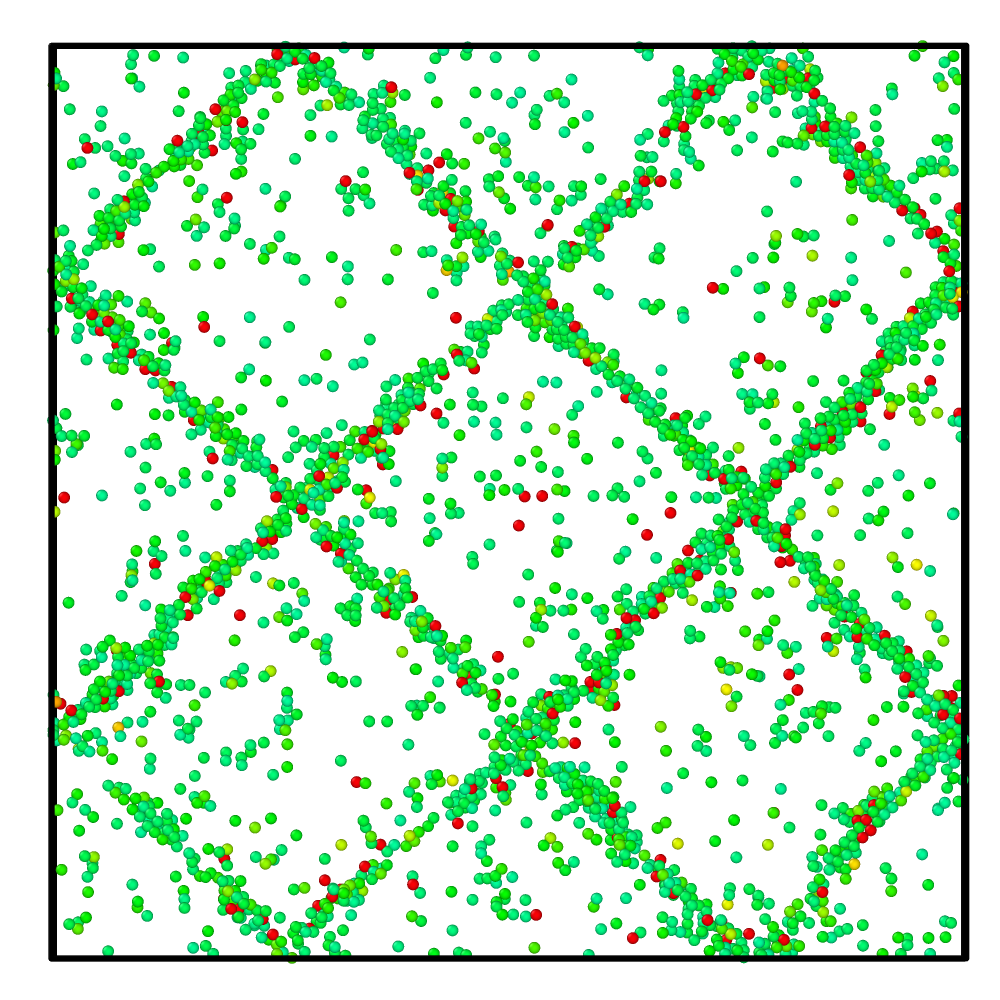}
\caption{GB structure at the temperature of 700 K. (a) Symmetrical tilt $\Sigma17(530)[001]$
GB. (b) $\Sigma3601(001)$ twist GB. The grain orientations are the
same as in Figure \ref{fig:GB17530}. The green color represents the
most distorted Al atoms with the centrosymmetry parameter above a
threshold value. The red color represents Mg atoms. The images have
been generated using OVITO\textcolor{blue}{{} }\citep{Stukowski2010a}. }
\label{fig:GBs_disorder}
\end{figure}

Mg was found to segregate to both GBs at all temperatures studied.
The images in Fig.~\ref{fig:GBs_disorder} illustrate the equilibrium
distributions of the Mg atoms along with the atomic disorder of the
GB structures at the temperature of 700 K.

\begin{figure}[ht]
\textbf{(a)} \includegraphics[height=0.4\textwidth]{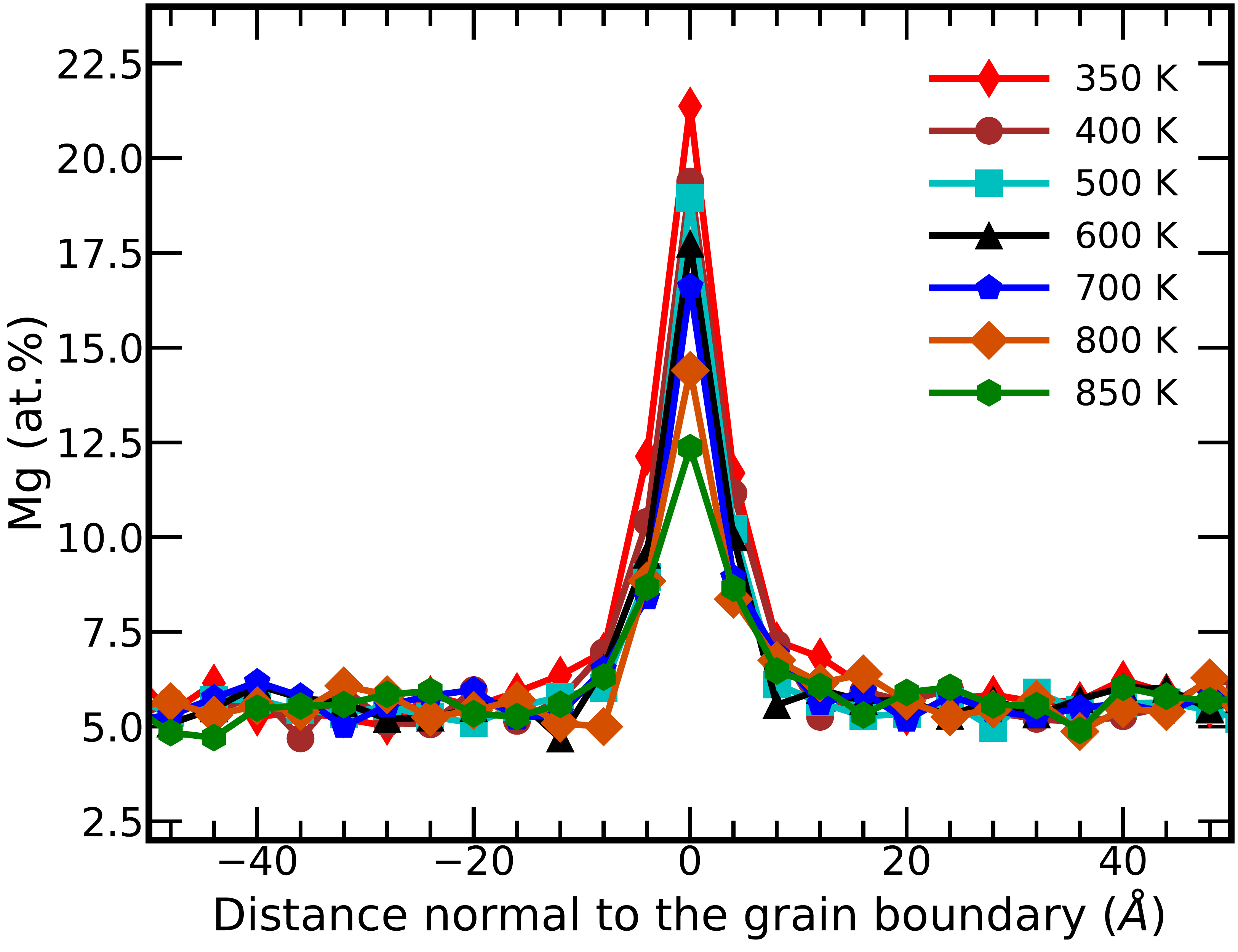}

\bigskip{}

\textbf{(b)} \includegraphics[height=0.4\textwidth]{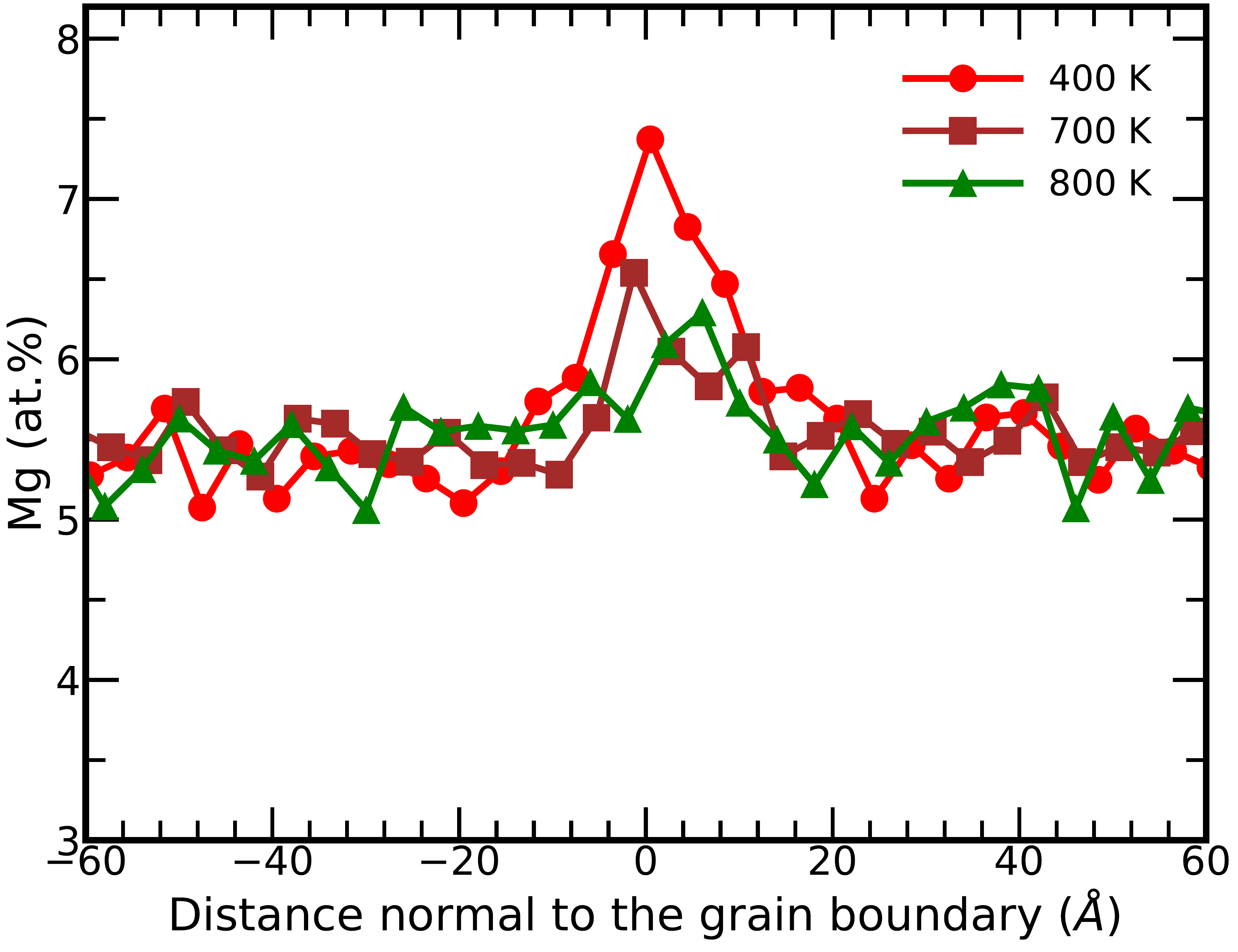}
\caption{Mg segregation profiles in (a) $\Sigma17(530)[001]$ tilt GB and (b)
$\Sigma3601(001)$ twist GB at several temperatures. The alloy composition
is Al-5.5at.\%Mg. }
\label{fig:GB_Segregation}
\end{figure}

Equilibrium segregation profiles were computed by averaging the atomic
fraction of Mg over thin layers parallel to the GB on either side
of its current position. The composition profiles displayed in Fig.~\ref{fig:GB_Segregation}
were averaged over multiple snapshots during the MD/MC simulations
after thermodynamic equilibration. The following features of the segregation
profiles are noted:
\begin{itemize}
\item Mg segregates to the high-angle GB much stronger than to the low-angle
GB.
\item The height of the segregation peak increases with decreasing temperature,
reaching about 21 at.\%Mg in the high-angle GB and about 7 at.\%Mg
in the low-angle GB at the lowest temperature tested.
\item At high temperatures approaching the melting point of the alloy ($>850$
K), the segregation profile of the high-angle GB significantly broadens,
suggesting that the boundary undergoes a premelting transformation. 
\end{itemize}
At temperatures between 860 and 870 K, the premelted high-angle GB
was observed to extend across the entire simulation block, transforming
it into the bulk liquid phase. Based on this observation, the solidus
temperature of the alloy was estimated to be $865\pm5$ K. This estimate
compares well with the equilibrium phase diagram obtained by independent
calculations in \citep{Mendelev:2009aa}. The low-angle GB did not
premelt and could be readily overheated above the solidus temperature,
keeping the dislocation network intact albeit with highly disordered
dislocation cores.

\begin{figure}[ht]
\textbf{(a)} \includegraphics[height=0.4\textwidth]{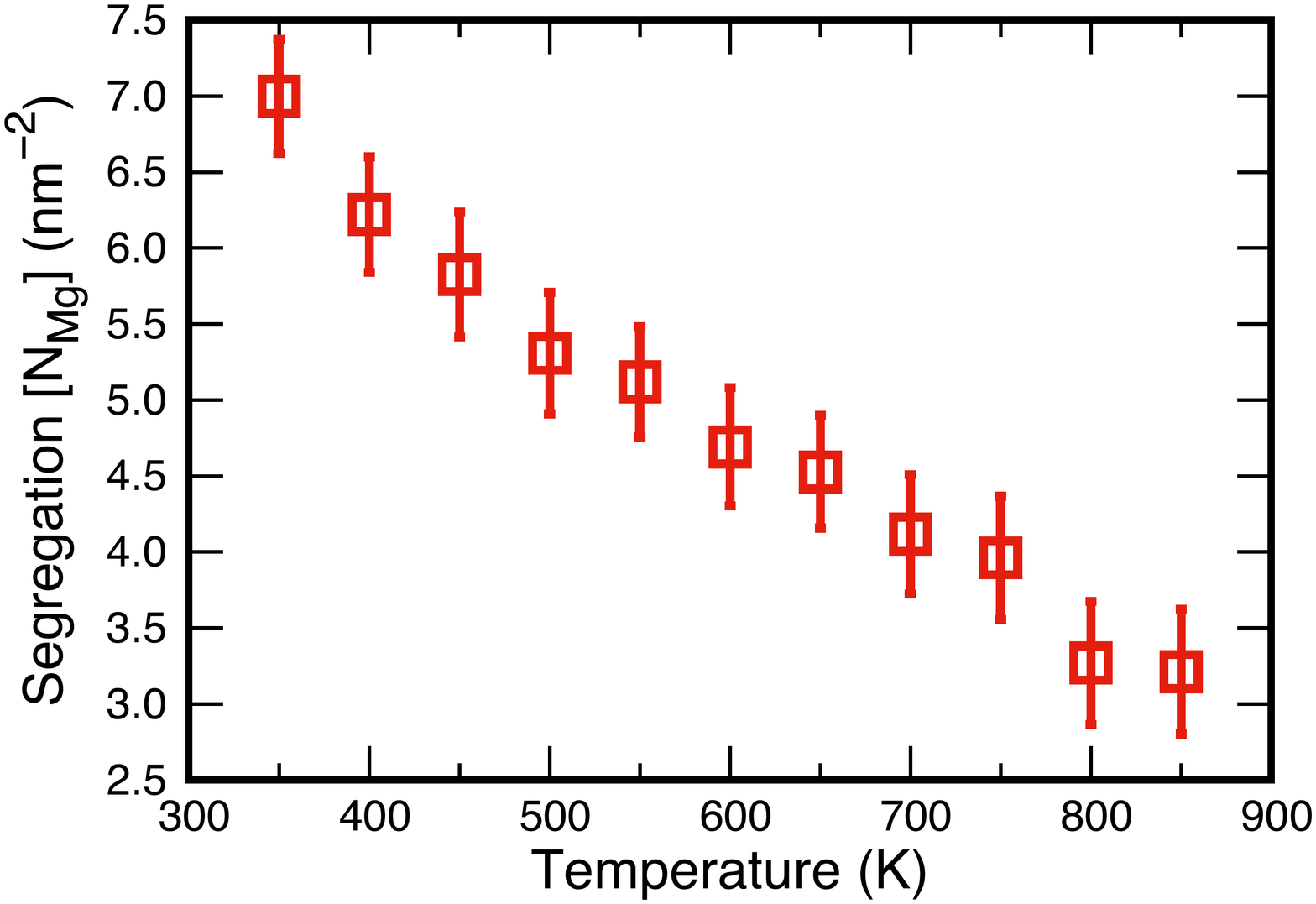}

\bigskip{}

\textbf{(b)} \includegraphics[height=0.4\textwidth]{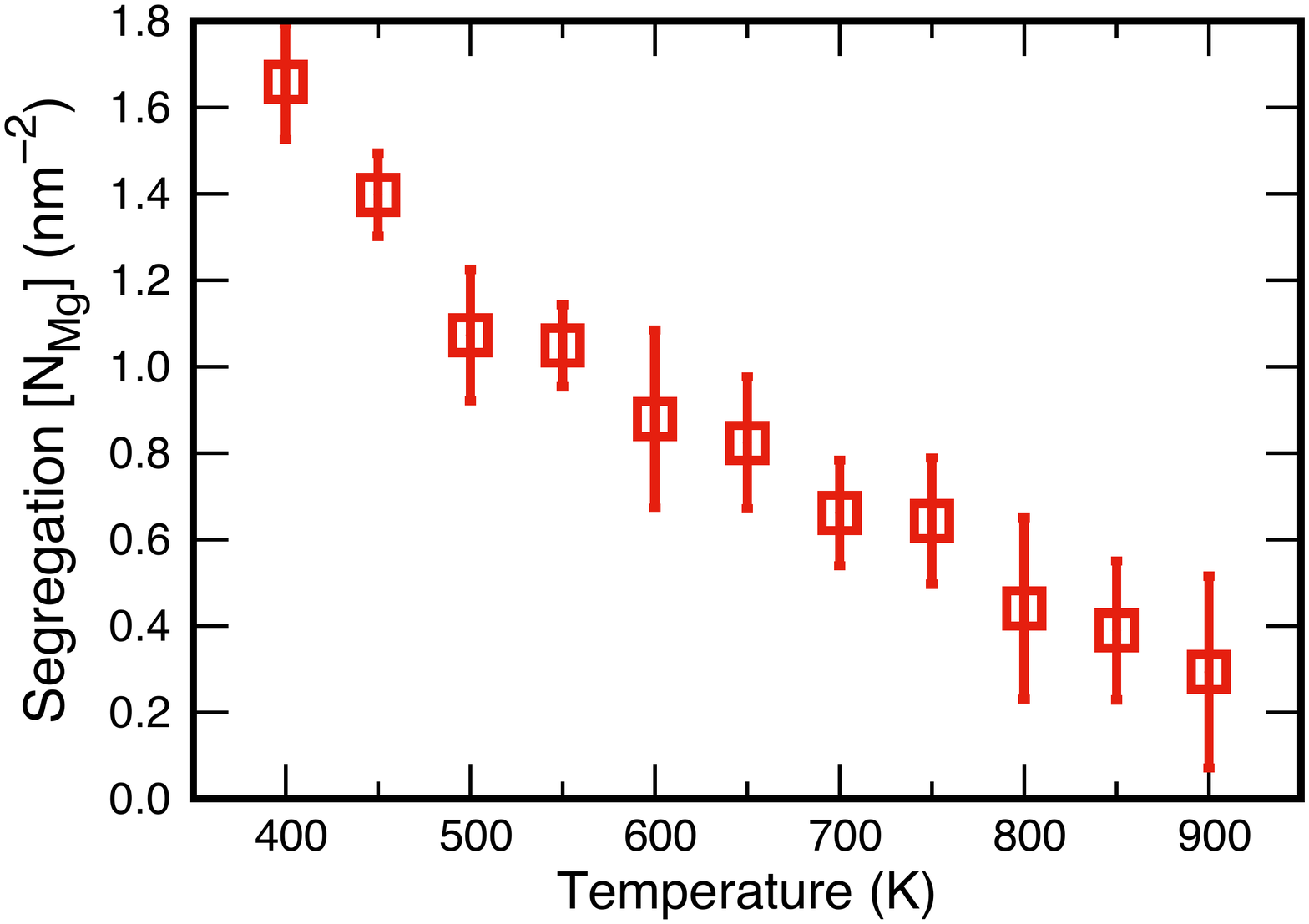}
\caption{Mg segregation in the Al-5.5at.\%Mg alloy as a function of temperature
for the (a) $\Sigma17(530)[001]$ tilt GB and (b) $\Sigma3601(001)$
GB. The error bars represent one standard deviation from averaging
over multiple snapshots.}
\label{fig:GB_Segregation-1}
\end{figure}

The amount of segregation was quantified by computing the excess number
of Mg atoms per unit GB area at a fixed total number of atoms:
\begin{equation}
[N_{\textnormal{Mg}}]=N_{\textnormal{Mg}}-N\dfrac{N_{\textnormal{Mg}}^{\prime}}{N^{\prime}}.\label{eq:1}
\end{equation}
Here, $N_{\textnormal{Mg}}$ and $N_{\textnormal{Mg}}^{\prime}$ are
the numbers of Mg atoms in two regions with and without the GB, respectively,
and $N$ and $N^{\prime}$ are the total numbers of Al and Mg atoms
in the respective regions. These regions were chosen to have the same
cross-sectional area parallel to the GB, and the excess $[N_{\textnormal{Mg}}]$
was normalized by this area. Accordingly, the units of $[N_{\textnormal{Mg}}]$
reported here are the number of excess Mg atoms per nanometer squared.
The average value and standard deviation of $[N_{\textnormal{Mg}}]$
were obtained by averaging over multiple snapshots generated during
the MC/MD simulations. Fig.~\ref{fig:GB_Segregation-1} shows the
amount of Mg segregation as a function of temperature. As expected
from the segregation profiles (cf.~Fig.~\ref{fig:GB_Segregation}),
$[N_{\textnormal{Mg}}]$ decreases with increasing temperature and
is much higher for the high-angle GB than for the low-angle GB. 

\begin{figure}[ht]
\textbf{(a)} \includegraphics[height=0.4\textwidth]{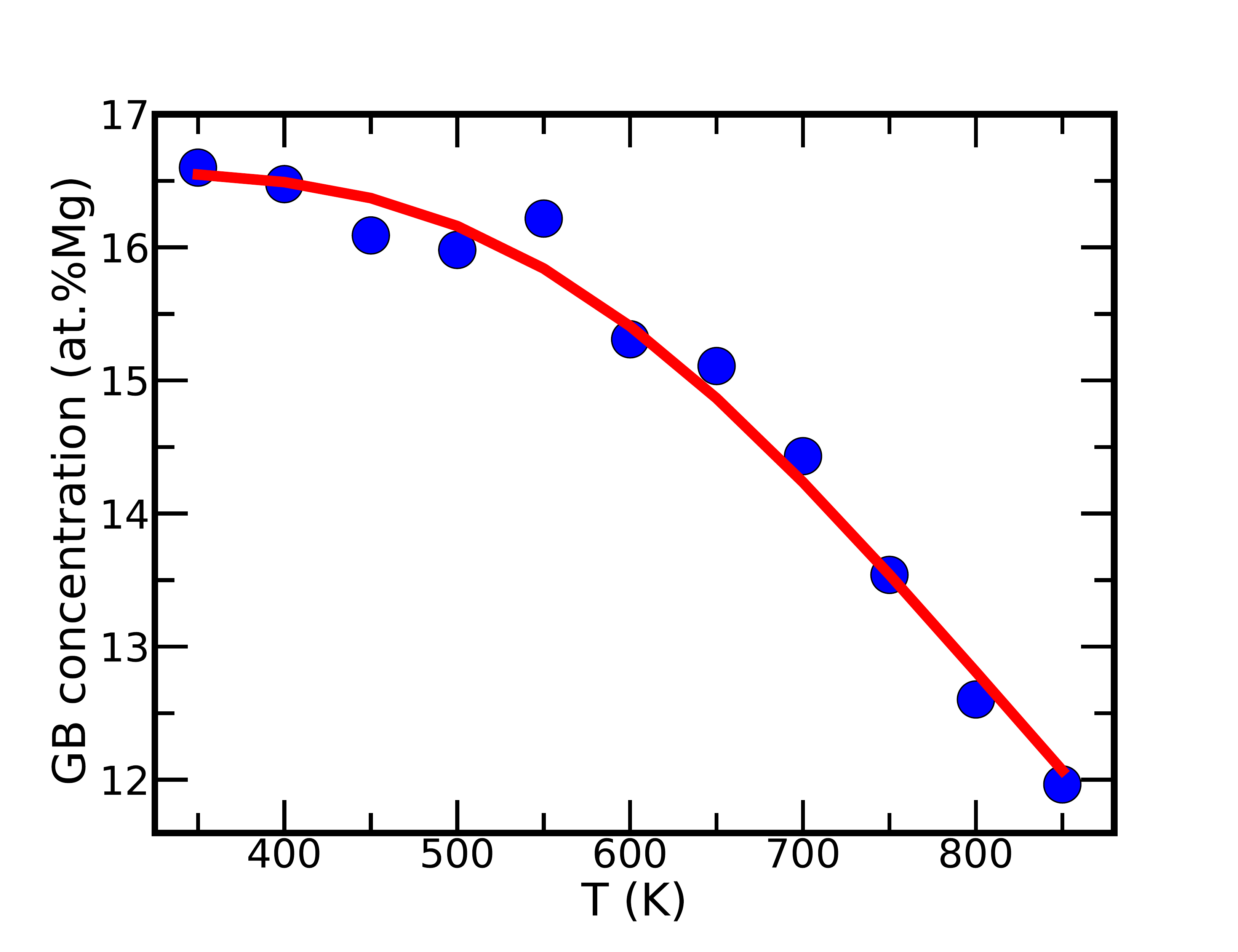}

\bigskip{}

\textbf{(b)} \includegraphics[height=0.4\textwidth]{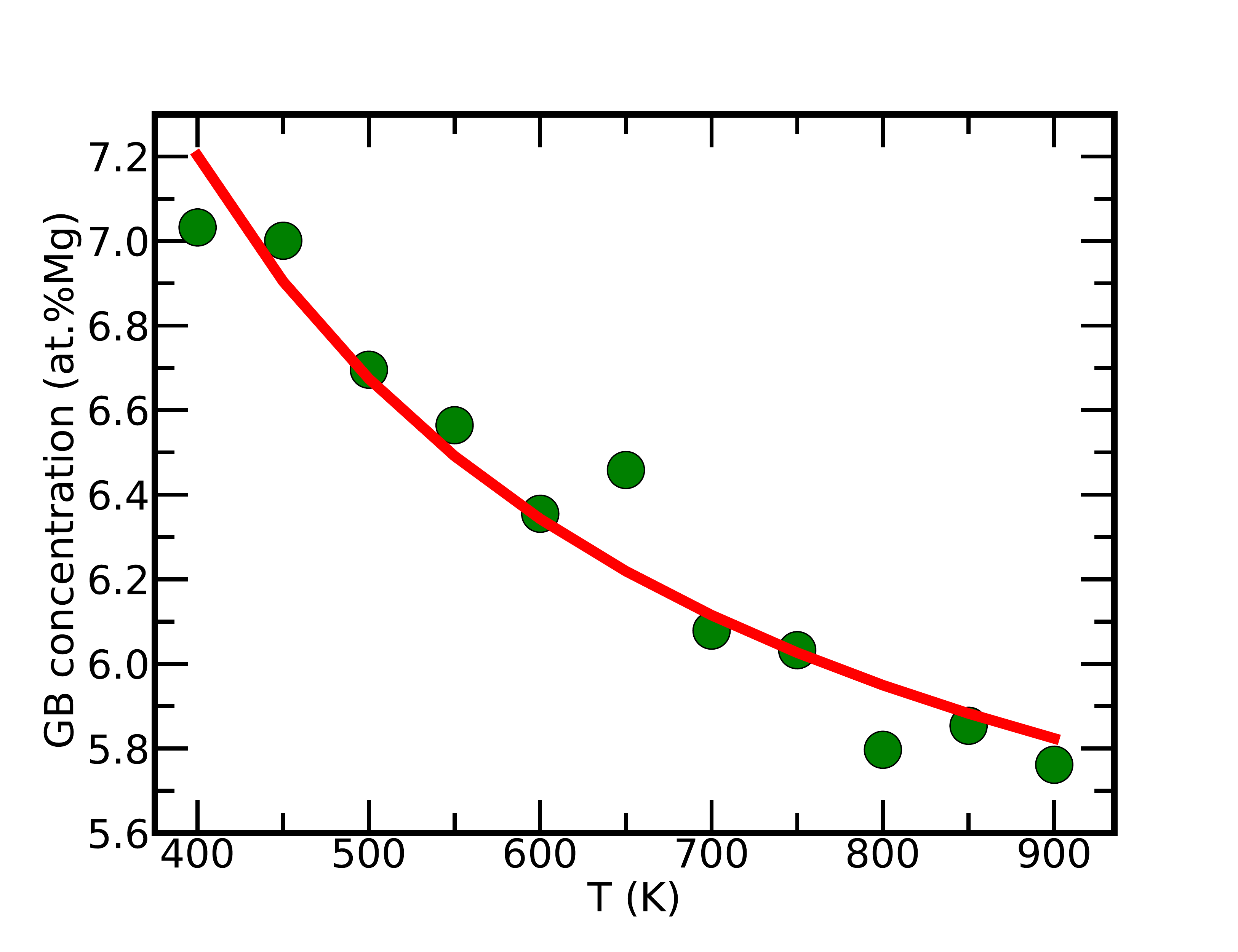}
\caption{Mg atomic fraction in the (a) $\Sigma17(530)[001]$ tilt GB and (b)
$\Sigma3601(001)$ GB as a function of temperature. The points represent
simulation results while the curves were obtained by fitting the Langmuir-McLean
model in Eq.(\ref{eq:2}).\label{fig:McLean}}
\label{fig:GB_Segregation-1-1}
\end{figure}

An alternative measure of the Mg segregation is the atomic fraction
$c_{GB}$ of Mg atoms in the GB computed by averaging over a layer
of the Gaussian width centered at the concentration peak (cf.~Fig.~\ref{fig:GB_Segregation}).
The GB concentrations obtained are expected to follow the modified
Langmuir-McLean segregation isotherm \citep{McLean}
\begin{equation}
\dfrac{c_{GB}}{\alpha-c_{GB}}=\dfrac{c}{1-c}\exp\left(-\dfrac{F_{s}}{kT}\right).\label{eq:2}
\end{equation}
Here, $c$ is the alloy composition (atomic fraction of Mg), $k$
is Boltzmann's constant, $F_{s}$ is the segregation free energy per
atom, and $\alpha$ is the fraction of GB sites filled by Mg atoms
when the segregation is fully saturated. $F_{s}$ represents the difference
between the free energies of Mg atoms inside the GB and in the grain
interiors. For both GBs, the temperature dependence of $c_{GB}$ could
be fitted by equation (\ref{eq:2}) reasonably well, see Fig.~\ref{fig:McLean},
with the values of $F_{s}$ and $\alpha$ listed in Table \ref{tab:McLean}.
For the low-angle GB, the quality of fit is somewhat lower because
$c_{GB}$ is significantly closer to $c$. The negative values of
$F_{s}$ indicate that the interaction between the Mg atoms and the
GBs is attractive. The absolute values of $F_{s}$ are also meaningful
and consistent with previous reports. For example, Mg segregation
energies in Al $\Sigma5\ [001]$ tilt and twist GBs were found to
be $-0.50$ eV and $-0.20$ eV, respectively \citep{Liu:1998aa}.
A more recent first-principles study of the Al $\Sigma5\ [001]$ tilt
boundary reports the Mg segregation energy of $-0.3$ eV \citep{Petrik:2018aa}.
For the Al $\Sigma11\ [311]$ tilt GB, first-principles calculations
predict the Mg segregation energies of $-0.02$ eV, $-0.070$ eV and
$-0.185$ eV for three different GB sites \citep{Liu:2005aa}. It
should be noted that the calculations in \citep{Liu:1998aa} utilized
a different interatomic potential, and that the values reported in
the literature represent the segregation energy, not free energy.
The free energy obtained here additionally includes the effects of
the vibrational and configurational entropies. Furthermore, GB structures
typically exhibit a diverse set of atomic environments, and thus a
wide spectrum of segregation energies. The values of $F_{s}$ reported
in Table \ref{tab:McLean} should be interpreted as representative
(effective) values. The saturation parameter $\alpha$ is understood
as the fraction of the GB sites with the largest magnitude of $F_{s}$.
Given these uncertainties, we consider our results to be in reasonable
agreement with the literature and consistent with the physical meaning
of segregation parameters.

\begin{figure}[ht]
\vspace{0cm}
 \centering \includegraphics[width=0.37\textwidth]{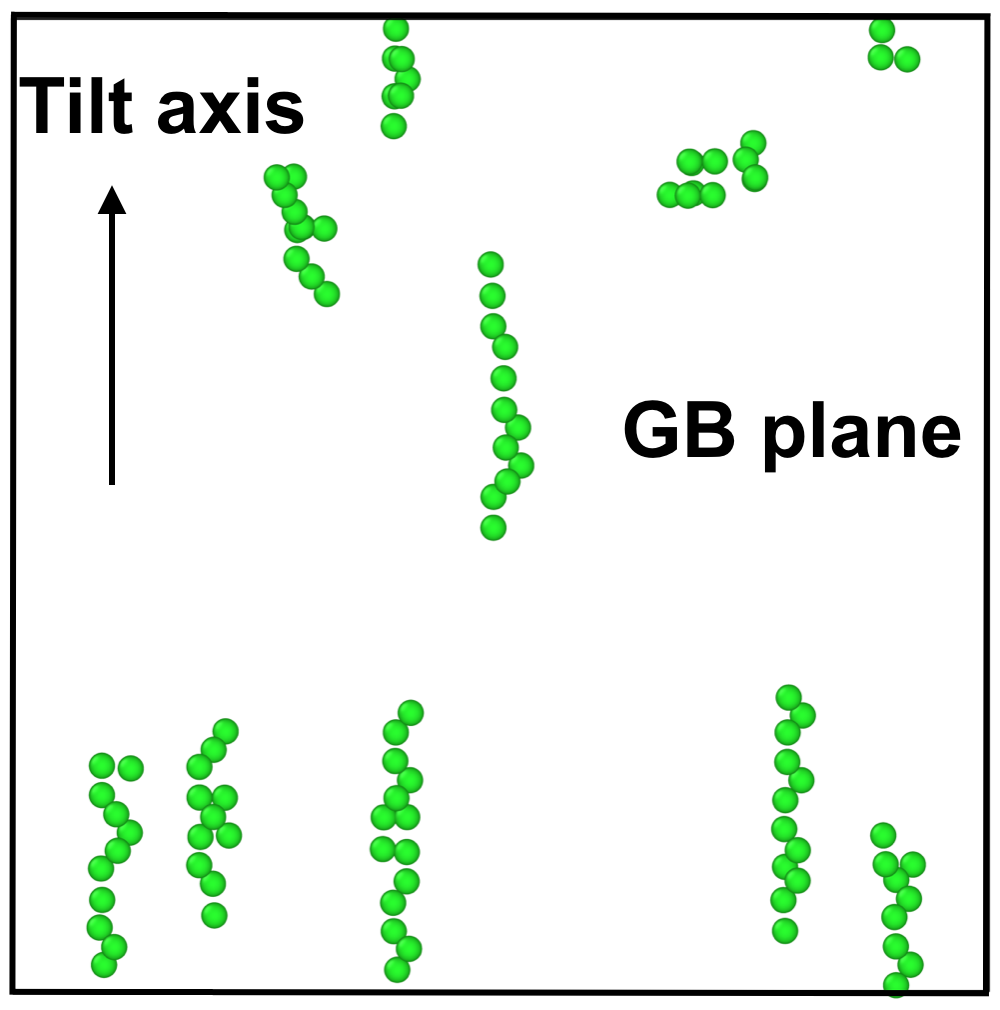} \caption{Mg clusters in the $\Sigma17(530)[001]$ tilt GB at 400 K. The GB
plane is parallel to the page. Only clusters containing 10 or more
atoms are shown for clarity. }
\label{fig:xDis-1}
\end{figure}

\begin{figure}[ht]
\textbf{(a)} \includegraphics[height=0.4\textwidth]{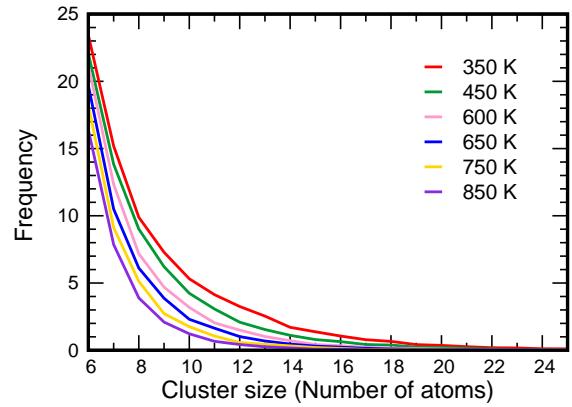}

\bigskip{}

\textbf{(b)} \includegraphics[height=0.4\textwidth]{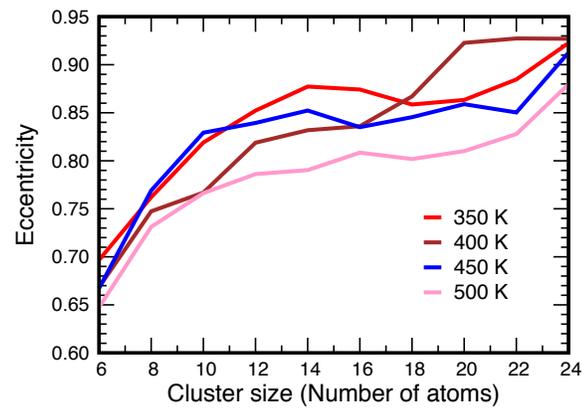}
\caption{Size and shape of Mg clusters in the $\Sigma17(530)[001]$ tilt GB
at selected temperatures. (a) Size distribution. The vertical axis
gives the number of clusters of a given size in the simulation block
averaged over multiple snapshots. (b) Eccentricity of the clusters,
given by Eq.(\ref{eq:eccentr}), plotted as a function of the cluster
size.}
\label{fig:Cluster_plots}
\end{figure}

A peculiar segregation feature was found in the high-angle GB. While
most of the Mg atoms were distributed in the GB in a random manner,
a tendency to form Mg clusters was observed, especially at low temperatures.
Cluster analysis was performed on statically relaxed snapshots using
the OVITO software \citep{Stukowski2010a}. An example of clusters
is shown in Fig.~\ref{fig:xDis-1}. To reveal the clustering effect
more clearly, only clusters containing 10 or more atoms are visualized.
Figure \ref{fig:Cluster_plots}a shows the cluster size distribution
at different temperatures (size being defined as the number of atoms
in the cluster). Only clusters containing 6 or more atoms are included
in the distribution. Since such clusters constitute a tiny fraction
of the entire cluster population in the GB, their contribution would
be nearly invisible if smaller clusters were included in the distribution.
At most temperatures, it was not unusual to see clusters containing
10 or more atoms. In fact, even clusters containing 30 to 40 atoms
were occasionally seen at low temperatures. It should be emphasized
that the clusters discussed here are not a static feature of the GB
structure. Instead, they behave as dynamic objects that randomly form
and dissolve during MD simulations, constantly changing their size,
shape and location by exchanging Mg atoms with each other and with
the bulk solution. The clustering of segregated atoms is a clear sign
of attractive solute-solute interactions inside the GB core. 

It should also be noted that the clusters shapes are significantly
elongated along the tilt axis. This elongation was quantified by the
eccentricity parameter
\begin{equation}
e=\sqrt{1-\dfrac{1}{2}\left(\dfrac{l_{y}}{l_{x}}\right)^{2}-\dfrac{1}{2}\left(\dfrac{l_{z}}{l_{x}}\right)^{2}},\label{eq:eccentr}
\end{equation}
where $l_{x}$ represents the cluster dimension along the tilt direction,
and $l_{y}$ and $l_{z}$ are the respective dimensions in the two
perpendicular directions. The eccentricity was calculated only when
the dimension along the tilt axis was longer than in the perpendicular
directions, and was assigned a zero value otherwise. As evident from
Fig.~\ref{fig:Cluster_plots}b, the cluster elongation tends to increase
(larger $e$) with the cluster size and decrease with temperature.
Large clusters containing 20 or more atoms looked almost like linear
chains. 

\subsection{Grain boundary diffusion}

\begin{figure}[ht]
\vspace{0cm}
 \centering \includegraphics[width=0.52\textwidth]{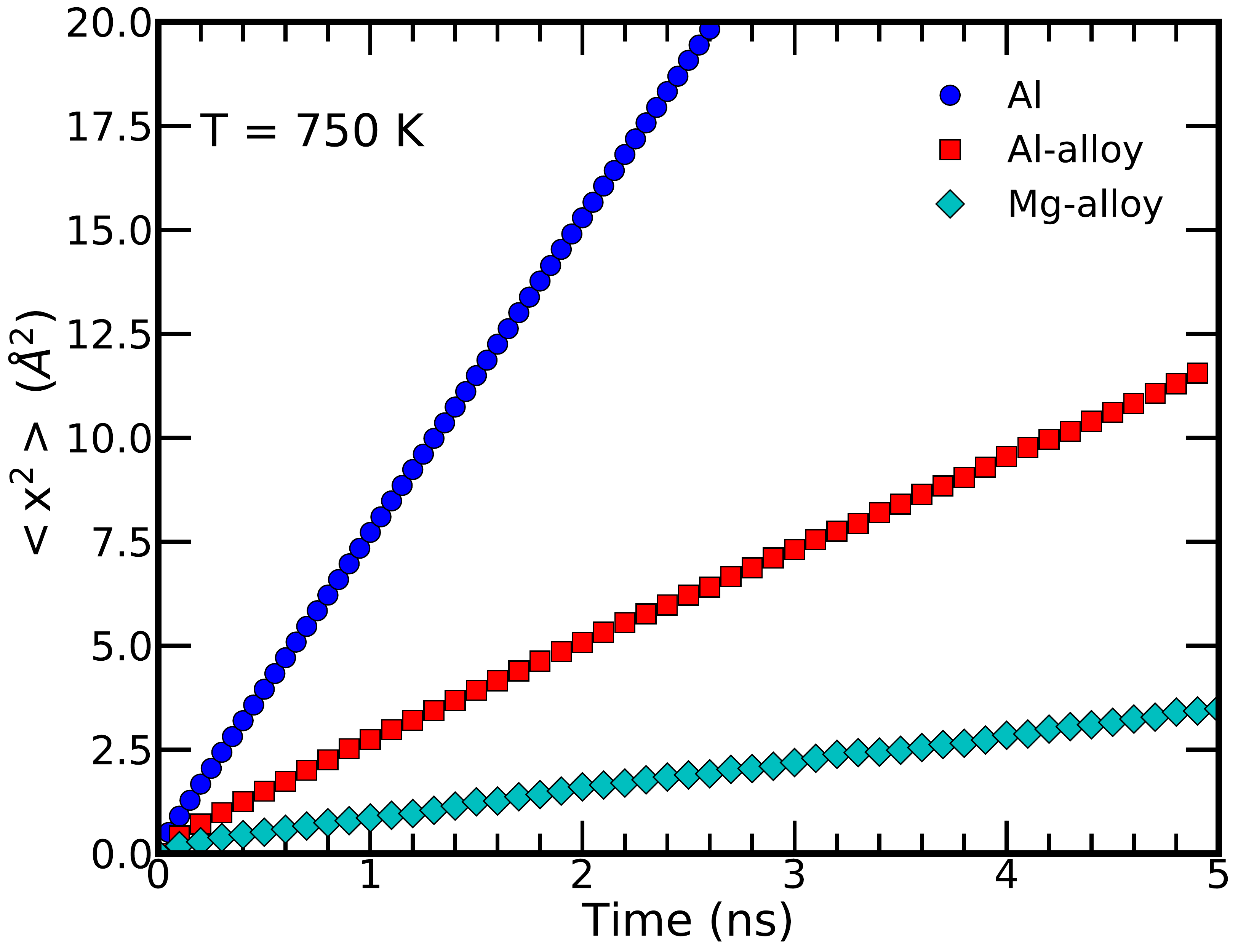}
\caption{Mean-square atomic displacement normal to the tilt axis versus time
in the $\Sigma17(530)[001]$ GB at the temperature of 750 K. The lines
represent GB self-diffusion in pure Al and GB diffusion of Al and
Mg in the Al-5.5at.\%Mg alloy. }
\label{fig:xDis}
\end{figure}

Figure \ref{fig:xDis} shows representative $\left\langle x^{2}\right\rangle $
versus time plots whose slopes were used for computing the GB diffusion
coefficients. The plots are fairly linear as expected from the Einstein
relation. The slopes of the plots indicate that Al GB self-diffusion
is faster than Al GB diffusion in the alloy, which in turn is faster
than Mg GB diffusion in the alloy. For the high-angle GB, this trend
holds at all temperatures studied here. In the low-angle GB, Al and
Mg diffuse at approximately the same rate, and both are slower in
comparison with Al self-diffusion. 

\begin{figure}[ht]
\vspace{0cm}
 \centering \includegraphics[angle=-90,width=0.9\textwidth]{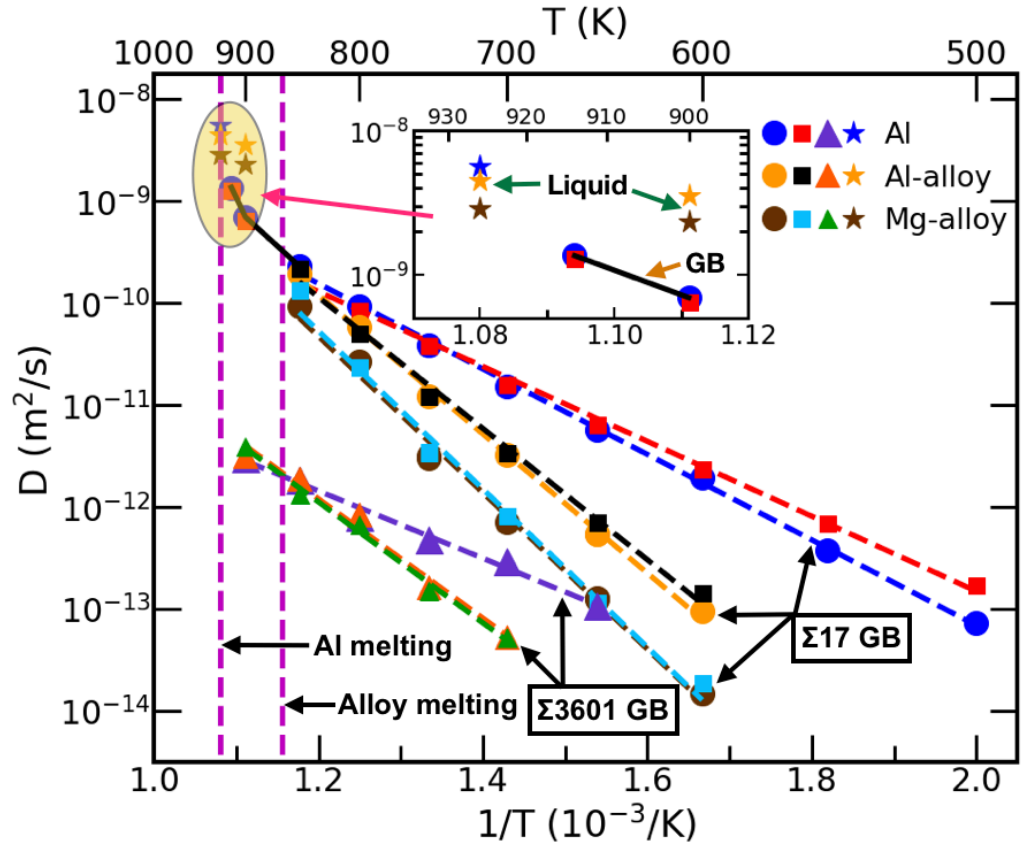}
\caption{Arrhenius diagram of GB diffusion coefficients (points) and their
liner fits (dashed lines). The square and circle symbols represent
diffusion parallel and normal to the tilt axis, respectively, in the
high-angle GB. The triangular symbols represent diffusion in the low-angle
GB. The inset is a zoom into the high-temperature region showing diffusion
in liquid Al and the liquid alloy (star symbols).}
\label{fig:Diffusion}
\end{figure}

The results of the diffusion calculations are summarized in the Arrhenius
diagram, $\log D$ versus $1/T$, shown in Fig.~\ref{fig:Diffusion}.
For the high-angle GB, the diffusion coefficients are reported separately
for both directions, parallel and perpendicular to the tilt axis.
Diffusion in the high-angle GB is several orders of magnitude faster
than diffusion in the low-angle GB at all temperatures. This behavior
is typical for metallic systems as reviewed in \citep{Kaur95,Mishin97e,Mishin99f}.
The diffusion coefficients closely follow the Arrhenius relation
\begin{equation}
D=D_{0}\exp\left(-\dfrac{E}{kT}\right)\label{eq:Arrhenius-eqn}
\end{equation}
at all temperatures below the solidus temperature. Note that Mg segregation
reduces or even eliminates the diffusion anisotropy in the high-angle
GB. In pure Al, diffusion along the tilt axis is faster than in the
direction normal to the tilt axis. This trend is general and was observed
in both experiments and previous simulations, for example in Cu and
Cu-Ag alloys \citep{Suzuki03a,Suzuki05a,Frolov2013a,Divinski2012}.
In the Al-Mg alloy, the anisotropy of Al GB diffusion is significantly
smaller in comparison with that of self-diffusion in pure Al. Furthermore,
GB diffusion of Mg is practically independent of the direction.

Table \ref{tab:The-activation-energy} summarizes the activation energies
$E$ and pre-exponential factors $D_{0}$ obtained by fitting Eq.(\ref{eq:Arrhenius-eqn})
to the simulation data. For the low-angle GB, the diffusivity follows
the Arrhenius relation even above the solidus temperature, which allowed
us to include one extra point (900 K) into the fit. Note that the
activation energies follow the trend $E_{\textrm{Al-Al}}<E_{\textrm{Al-Alloy}}<E_{\textrm{Mg-Alloy}}$,
suggesting that the observed retardation of GB diffusion by Mg segregation
is primarily caused by increase in the activation energy. This is
also evident from the converging behavior of the Arrhenius lines in
Fig.~\ref{fig:xDis}, leading to very similar diffusion coefficients
of Al and Mg close to the melting point.

In pure Al, the self-diffusivity in the high-angle GB was also computed
at two additional temperatures (900 and 914 K) lying above the alloy
solidus temperature but below the Al melting point (926 K). At these
temperatures, the boundary develops a highly disordered atomic structure
similar to a liquid layer. Accordingly, the GB diffusion coefficient
shows a significant upward deviation from the Arrhenius behavior and
approaches the self-diffusion coefficient in liquid Al (see inset
in Fig.~\ref{fig:xDis}). A similar behavior was previously observed
in the same $\Sigma17$ GB in Cu \citep{Suzuki05a}. It is interesting
to note that Al diffuses in the liquid alloy somewhat slower than
in pure Al, and Mg diffused even slower. This trend mimics the similar
behavior of GB diffusion, suggesting that the underlying cause is
the nature of atomic interactions in the Al-Mg system rather than
details of the GB structures.

\section{Discussion}

Atomistic simulations of GB structure, solute segregation and GB diffusion
are computationally expensive and have only been preformed for a small
number of GBs in a few binary systems. Systematic investigations covering
a wide range of temperatures all the way to the melting point are
especially demanding. For this reason, only two GBs have been studied
in the present work. As such, we selected two boundaries belonging
to very different classes: a low-angle GB, which essentially represents
a dislocation network, and a high-angle GB with a structurally homogeneous
core. Although each boundary is characterized by specific set of crystallographic
parameters, many of the conclusions of this work are generic and should
be valid for all low-angle and all high-angle GBs, respectively. In
particular, the fact that diffusion in the low-angle GB is slower
and is characterized by a larger activation energy in comparison with
the high-angle GB, is consistent with the existing body of experimental
data for many other alloy systems \citep{Kaur95}. The retardation
of Al diffusion by the presence of Mg atoms was found in both low-angle
and high-angle GBs, as well as in the bulk liquid phase, which strongly
suggests that this is a generic effect. It should also be noted that
at most temperatures studied in this work, the high-angle GB was found
to be structurally disordered. In fact, at high enough temperatures
it becomes a liquid-like layer. Under such conditions, the specific
bicrystallography of this boundary is unimportant and it can be considered
a ``generic'' high-angle GB. 

There are several findings in this paper whose explanation requires
furthers research. One of them is the observation of the strongly
elongated Mg clusters (atomic chains) in the high-angle GB. We hypothesize
that such clusters, as well as other possible chemical heterogeneities
in segregated Al GBs, can serve as precursors of Al-Mg intermetallic
compounds during their nucleation in oversaturated alloys. The clustering
trend also suggests that the GB solution has a miscibility gap. While
this line of inquiry was not pursued in this work, it seems quite
possible that Al-Mg GBs can exhibit 2D phases and phase transformations
among them \citep{Frolov:2015ab,Mishin:2019aa}. Furthermore, it is
likely that the Mg clusters act as traps for diffusion of Mg atoms,
vacancies and interstitials. This would explain the relatively show
GB diffusion rate of Mg. However, further work is required to better
understand the underlying atomic mechanisms.

Although the GB diffusivities reported here cannot be compared with
experiments, the Mg GB segregation in Al has been studied by several
experimental techniques, including atom probe tomography (APT). The
experiments show that Mg strongly segregates to Al GBs in most cases
\citep{Guillot:2010aa,Liu:1998aa,Liu:2005aa,Lee:2016aa,Pun:2017aa,Devaraj:2019aa,Rahman:2016aa,Kazemi:2019aa}.
However, deviation from this trend were also reported in the literature.
For example, recent APT studies of Mg distribution after severe plastic
deformation \citep{Xue:2019aa,Sauvage:2019aa} revealed Mg-depleted
zones near GBs. These zones are explained \citep{Xue:2019aa} by inhomogeneous
nature of the deformation process, namely, by the interaction of Mg
atoms with moving dislocations in micro-deformation bands in the deformed
microstructure. This highly non-equilibrium effect does not contradict
the observation of equilibrium Mg segregation in this work as well
as in previous reports.

On the simulation side, Mg GB segregation in nanocrystalline Al-Mg
was recently studied by the lattice Monte Carlo (LMC) method \citep{Pun:2017aa}.
This method is different from the potential-based off-lattice Monte
Carlo simulations reported in this paper. In LMC simulations, the
lattice remains rigid and the interaction parameters are fitted to
experimental information within the regular solution approximation.
GBs are defined as regions with modified values of the interaction
parameters. Despite these differences, the LMC results are consistent
with our work. For example, the segregation isotherm at 200$^{\lyxmathsym{\textdegree}}$C
and the alloy composition of about 5 at.\%Mg (Figure 7a in \citep{Pun:2017aa})
predicts GB concentration of about 30 at.\%Mg. Our simulations give
the concentration of about 22 at.\%Mg at 350 K (Fig.~\ref{fig:GB_Segregation}a).
Furthermore, the interaction of Mg atoms with GBs was recently studied
by first-principles calculations \citep{Petrik:2018aa} using the
$\Sigma5$ (201){[}001{]} symmetrical tilt boundary as a model. The
calculations confirm a negative segregation energy of Mg driving GB
segregation. At the temperature of 550 K, the peak Mg concentration
in this boundary was found to be about 32 at.\%Mg. Thus, calculations
by different methods for different high-angle GBs in Al predict the
segregation levels of Mg consistent with the present work. This agreement
is reassuring and suggests that the results reported here reflect
the generic nature of the Mg interaction with Al GBs.

\section{Conclusions\label{sec:Conclusions}}

We have studied GB segregation and GB diffusion in the Al-Mg system
by atomistic computer simulations combining MD and MC methods. A typical
Al-5.5at.\%Mg alloy and two representative (high-angle and low-angle)
GBs were chosen as models. The conclusions can be summarized as follows:
\begin{itemize}
\item In agreement with previous reports, Mg strongly segregates to high-angle
GBs and, to a lesser extent, to low-angle GBs composed of dislocations.
At low temperatures, such as 350 K, the local chemical composition
in GBs can exceed 20 at.\%Mg.
\item The amount of GB segregation increases with decreasing temperature.
The effective free energy of GB segregation is estimated to be about
$-0.28$ eV/atom for the high-angle GB studied here and much smaller
($\sim-0.01$ eV/atom) for the low-angle GB.
\item Distribution of the segregated Mg atoms over a GB is highly non-uniform.
In the high-angle tilt GB, the Mg atoms tend to form clusters containing
10 to 30 atoms, especially at low temperatures. Such clusters are
elongated parallel to the tilt axis and are similar to linear atomic
chains. 
\item At high temperatures approaching the solidus line, the high-angle
GB studied here exhibits a premelting behavior by developing a highly
disordered, liquid-like structure. By contrast, the low-angle GB does
not premelt and can be overheated past the solidus line. While the
individual dislocations do become disordered, the dislocation network
itself remains intact, demonstrating an extraordinary thermal stability.
\item Mg segregation strongly affects the rate of GB diffusion in Al-Mg
alloys. Mg GB diffusion is slower than Al GB self-diffusion in pure
Al. Furthermore, Mg segregation slows down the GB diffusion of Al
itself. This diffusion retardation could be responsible for the microstructure
stability in Al-Mg alloys. 
\item The diffusion retardation effect caused by the Mg segregation is primarily
due to the significant (about a factor of two) increase in the activation
energy of GB diffusion (Table \ref{tab:The-activation-energy}). 
\item Mg segregation reduces the anisotropy of GB diffusion.
\item Mg diffusion in high-angle GBs is several orders of magnitude faster
than diffusion in low-angle GBs at the same temperature.
\end{itemize}
In the absence of experimental data, the GB diffusion coefficients
obtained in this work can provide useful reference information for
further investigations of Al-Mg alloys. GB diffusion coefficients
appear as input material parameters in many models describing processes
such precipitation aging, solute drag, and micro-creep to name a few.

\vspace{0.15in}

\textbf{Acknowledgement:} R.~K.~K.~and Y.~M.~were supported by
the National Science Foundation, Division of Materials Research, under
Award No.1708314.

%\bibliographystyle{/Users/ymishin/YURI/Bibliography/Phys_Rev_Style}
%\bibliography{/Users/ymishin/YURI/Bibliography/literat}

\newpage\clearpage{}

\begin{table}
\noindent \begin{centering}
\begin{tabular}{|c|c|c|c|}
\hline 
Grain boundary & $F_{s}$ (eV) & $\alpha$ & $R^{2}$\tabularnewline
\hline 
\hline 
$\Sigma17(530)[001]$ tilt & $-0.281\pm0.004$ & $0.166\pm0.001$ & $98.39\%$\tabularnewline
\hline 
$\Sigma3601(001)$ twist & $-0.014\pm0.001$ & $0.891\pm0.021$ & $93.88\%$\tabularnewline
\hline 
\end{tabular}
\par\end{centering}
\caption{Segregation free energy and the fraction of available segregation
sites extracted from the simulation results. The last column reports
the $R^{2}$ coefficient of determination characterizing the qualify
of fit by the Langmuir-McLean model in Eq.(\ref{eq:2}).\label{tab:McLean}}
\end{table}

\begin{table}
\begin{tabular}{lcccccc}
\hline 
 & Direction & Al in pure Al &  & Al in alloy &  & Mg in alloy\tabularnewline
\hline 
\multicolumn{7}{c}{$\Sigma17(530)[001]$ GB}\tabularnewline
\multirow{2}{*}{$E$ (eV)} & $\parallel$ tilt axis & $0.73\pm0.02$ &  & $1.22\pm0.05$ &  & $1.52\pm0.08$\tabularnewline
 & $\bot$ tilt axis & $0.83\pm0.01$ &  & $1.27\pm0.03$ &  & $1.54\pm0.06$\tabularnewline
\multirow{2}{*}{$D_{0}$ (m$^{2}$/s)} & $\parallel$ tilt axis & $\left(3.33_{-0.90}^{+1.23}\right)\times10^{-6}$ &  & $\left(2.60_{-1.48}^{+3.45}\right)\times10^{-3}$ &  & $\left(8.48_{-6.06}^{+21.26}\right)\times10^{-2}$\tabularnewline
 & $\bot$ tilt axis & $\left(1.57_{-0.28}^{+0.34}\right)\times10^{-5}$ &  & $\left(5.38_{-2.16}^{+3.60}\right)\times10^{-3}$ &  & $\left(1.12_{-0.72}^{+1.99}\right)\times10^{-1}$\tabularnewline
\multicolumn{7}{c}{$\Sigma3601(001)$ GB}\tabularnewline
$E$ (eV) & $\bot$ twist axis & $0.66\pm0.04$ &  & $1.16\pm0.09$ &  & $1.18\pm0.06$\tabularnewline
$D_{0}$ (m$^{2}$/s) & $\bot$ twist axis & $\left(1.33_{-0.55}^{+0.93}\right)\times10^{-8}$ &  & $\left(1.27_{-0.94}^{+3.56}\right)\times10^{-5}$ &  & $\left(1.47_{-0.86}^{+2.08}\right)\times10^{-5}$\tabularnewline
\hline 
\end{tabular}

\caption{The activation energy $E$ and pre-exponential factor $D_{0}$ for
GB diffusion in pure Al and in the Al-Mg alloy.\label{tab:The-activation-energy}}
\end{table}

\newpage\clearpage{}

\end{document}